\documentclass[twocolumn,pre,showpacs,superscriptaddress]{revtex4}
\usepackage{graphicx}%
\usepackage{psfrag}
\usepackage{dcolumn}
\usepackage{amsmath}
\usepackage{amssymb}
\usepackage{latexsym}
\usepackage{hyperref} 
\usepackage{booktabs}
\begin{document}

\def\bea{\begin{eqnarray}}
\def\eea{\end{eqnarray}}
\def\beq{\begin{equation}}
\def\eeq{\end{equation}}
\def\f{\frac}
\def\k{\kappa}
\def\e{\epsilon}
\def\ve{\varepsilon}
\def\be{\beta}
\def\D{\Delta}
\def\h{\theta}
\def\t{\tau}
\def\a{\alpha}

\def\cDa{{\cal D}[X]}
\def\cD{{\cal D}[x]}
\def\cL{{\cal L}}
\def\cLo{{\cal L}_0}
\def\cLa{{\cal L}_1}

\def\Re{{\rm Re}}
\def\sj{\sum_{j=1}^2}
\def\rk{\rho^{ (k) }}
\def\rek{\rho^{ (1) }}
\def\cek{C^{ (1) }}
\def\rz{\rho^{ (0) }}
\def\rt{\rho^{ (2) }}
\def\rtb{\bar \rho^{ (2) }}
\def\trk{\tilde\rho^{ (k) }}
\def\trek{\tilde\rho^{ (1) }}
\def\trz{\tilde\rho^{ (0) }}
\def\trt{\tilde\rho^{ (2) }}
\def\r{\rho}
\def\tD{\tilde {D}}

\def\s{\sigma}
\def\kb{k_B}
\def\la{\langle}
\def\ra{\rangle}
\def\nn{\nonumber}
\def\up{\uparrow}
\def\dn{\downarrow}
\def\S{\Sigma}
\def\dg{\dagger}
\def\d{\delta}
\def\p{\partial}
\def\l{\lambda}
\def\L{\Lambda}
\def\G{\Gamma}
\def\o{\Omega}
\def\w{\omega}
\def\g{\gamma}

\def\noi{\noindent}
\def\a{\alpha}
\def\d{\delta}
\def\p{\partial} 

\def\la{\langle}
\def\ra{\rangle}
\def\e{\epsilon}
\def\n{\eta}
\def\g{\gamma}
\def\break#1{\pagebreak \vspace*{#1}}
\def\hf{\frac{1}{2}}

\title{Stochastic thermodynamics of active Brownian particles} 
\author{Chandrima Ganguly}
\affiliation{Indian Institute of Technology Hyderabad,
Yeddumailaram 502205, Andhra Pradesh, India
}
\author{Debasish Chaudhuri}
\email{debc@iith.ac.in}
\affiliation{Indian Institute of Technology Hyderabad,
Yeddumailaram 502205, Andhra Pradesh, India
}

\date{\today}

\begin{abstract}
Examples of self propulsion in strongly fluctuating environment is abound in nature, e.g.,  molecular motors and pumps operating in living cells. Starting from Langevin equation of motion, we develop a stochastic thermodynamic description of non-interacting self propelled particles using simple models of velocity dependent forces. 
We derive fluctuation theorems for entropy production and a modified fluctuation dissipation relation, characterizing the linear response  at non-equilibrium steady states. We study these notions in a simple model of molecular motors, and in the Rayleigh-Helmholtz  and  energy-depot model of self propelled particles. 
\end{abstract}

\maketitle

\section{Introduction}
Living systems are by definition open and active, staying out of equilibrium by consuming and subsequently dissipating energy, thereby generating forces and motion. Subcellular components, e.g.,  motor proteins, cytoskeletal filaments etc. operate in a stochastic environment, where fluctuations arise from thermal motion and, in many cases,  
chemical reactions~\cite{Alberts}. 
In contrast to conventional Brownian motion where the forces acting on a particle are entirely due to external sources, the {\em active} Brownian particles can generate their own forces~\cite{Vicsek2005} utilizing chemical energies~\cite{Alberts}. 

Traditional thermodynamics in terms of average quantities does not provide satisfactory description of small assembly of colloidal particles, or nano-materials due to the presence of strong thermal fluctuations. 
In the last two decades, a theoretical framework has emerged that allows several exact relations for distributions of fluctuating quantities like work,  heat and entropy characterizing 
individual trajectories of the particles~\cite{Seifert2012,Jarzynski2011,Baiesi2009a,Baiesi2010,Hummer2010,Saha2009, Imparato2006, Kurchan2007,Narayan2004,Crooks1999, Lebowitz1999,Jarzynski1997,Gallavotti1995,Evans1993}. 
At non-equilibrium steady states (NESS) entropy $\s$ is continually produced, with its probability distribution obeying~\cite{Seifert2005,Gallavotti1995,Evans1993}
${P(\s)}/{P(-\s)}=\exp(\s/\kb)$.
This is known as the detailed fluctuation theorem (DFT) and was first observed in simulations of sheared liquids~\cite{Evans1993} and later derived using chaotic~\cite{Gallavotti1995} and stochastic  dynamics~\cite{Kurchan2007,Lebowitz1999}. For asymptotic steady states the above relation is obeyed with $\s=\D s_m$,  $\D s_m$ being the 
change in entropy of the medium alone. If one considers the  stochastic change in system entropy $\D s$ as well,  $\s=\D s_{tot}=\D s + \D s_m$ signifying the total entropy change, the DFT remains
valid even for finite time measurements~\cite{Seifert2005}. 
Further, $\D s_{tot}$ obeys an integral fluctuation theorem $\la \exp(-\D s_{tot}/\kb) \ra= 1$ where $\la \dots \ra$ denotes non-equilibrium average over stochastic paths.
This is closely related to the Jarzynski equation, that expresses equilibrium free energy difference in terms of non-equilibrium work done~\cite{Jarzynski1997,Crooks1999}.  
These fluctuation theorems were verified in experiments on colloids~\cite{Wang2002, Blickle2006, Speck2007},
 and granular matter~\cite{Joubaud2012}, and successfully used to find out the free energy landscape of 
 RNA~\cite{Liphardt2002,Collin2005}. 
  Fluctuation theorems were also derived for the flashing ratchet~\cite{Lacoste2011,Lacoste2009}, and other detailed models of molecular motors and enzymes~\cite{Seifert2011}.
 However, given the complexity of living systems it may not  always be possible to identify and model all the chemical processes and  mechano-chemical coupling responsible for autonomous force generation. 
 Recently, the DFT was applied to measure the torque generation by a rotory motor F$_1$ATPase from its fluctuating trajectories~\cite{Hayashi2010}. 
 This idea may be extended to other types of molecular motors to measure  autonomous force or torque generation from their stochastic trajectories~\cite{Hayashi2012}.

 Response in equilibrium states is characterized by the fluctuation-dissipation theorem (FDT), and the ratio of correlation and response is often interpreted
 as effective temperature of systems at NESS~\cite{Kikuchi2009,Wang2011}. Recent theoretical work derived several forms of 
modified fluctuation-dissipation relations (MFDR) characterizing linear response at NESS and established additive correction to FDT  due to the presence of non-zero 
steady state currents\cite{Cugliandolo1994,Speck2006,Baiesi2009,Prost2009,Seifert2010,Verley2011,Chaudhuri2012}, thus showing that phenomenological characterization of active
processes by effective temperatures is not consistent.
Some of the theoretical predictions were verified experimentally~\cite{Blickle2007,Gomez-Solano2009}.

 In this paper, starting from the equation of motion for  a self propelled particle (SPP) immersed in a Langevin heat bath, we develop its stochastic thermodynamic description. 
 We assume that the self-propulsion force is velocity dependent, however, the details of the 
 propulsion mechanism is not specified to begin with. Using this assumption, we derive 
 energy balance relation and fluctuation theorems involving entropy production. 
 In particular, we identify the contributions in entropy production due to
 self-propulsion and its coupling to external drive. 
 We also obtain a modified fluctuation-dissipation relation characterising the linear response at steady states of an SPP. 
Finally, we use our theoretical development to study some specific model systems that utilises velocity dependent forces: 
a simple model of molecular motors obeying a linear force-velocity relation,  and models utilising non-linear velocity dependent  forces 
as self propulsion mechanism, namely, the Rayleigh-Helmholtz model and the energy-depot model.     
\section{Langevin equation and the laws of thermodynamics}
To develop the notions of stochastic thermodynamics of  SPP systems, let us focus on one dimension (1d), for simplicity. Simplest models of SPPs, like that  the
Rayleigh-Helmholtz model or the energy-depot model, use velocity dependent autonomous force $F(v)$ to model self-propulsion. The Langevin equation for the motion of each particle evolving in the presence of a time dependent external force $f(t)$ has the form
\bea
\dot x &=& v \nn\\
\dot v &=&  -\g v + \eta + F(v) - \f{\p U(x)}{\p x} + f(t) 
\label{lange}
\eea 
where $-\g v$ is viscous dissipation, $\eta$ is Gaussian white noise characterized by $\la \eta(t) \ra =0$, $\la \eta(t) \eta (t')\ra = 2  D_0 \d(t-t')$ with $D_0=\g \kb T$, 
$U(x)$ is a conservative external potential.  
We use particle mass $m=1$, unless otherwise specified.

\subsection{First law}
Multiplying the above equation by velocity $v$ and integrating over a small time interval $\t$ one obtains the first law~\cite{Sekimoto1998}
\bea
\D E =  \D W + \D q 
\label{1st}
\eea 
where $ E= (1/2) v^2 + U(x)$, $\D E$ denoting the change in internal energy, $\D W = \int^\t dt\, v. f(t)$ the work done on the SPP by external force, and the total heat flow
$\D q = \D Q + \D Q_m$ has two components, $ \D Q= \int^\t dt\, v. (-\g v + \eta) $  the energy flow from the heat bath,
$\D Q_m = \int^\t dt\, v. F(v)$  the energy flow from the internal {\em motor} degrees of freedom of the SPP.  
The presence of energy flow from internal motor differentiates SPPs from passive Brownian particles~\cite{Zimmermann2012a}. 

\subsection{Fluctuation theorem: connection with second law }
Consider the time evolution of the system from $t=0$ to $\tau$ through a path defined by $X=\{x(t), v(t), f(t) \}$. 
The probability of this path is given by (see Appendix-\ref{ap_fpath}) 
\bea
{\cal P}_+ 
&=& {\cal N} \d(\dot x - v) \exp \left[ -\hf \int_0^\t dt  \f{\p g(v)} {\p v} \right] \nn\\
&\times& \exp\left[{-\frac{1}{4D_0} \int_0^\tau  dt \left( \dot v - g(v) + \f{\p U}{\p x} -f(t) \right)^2}\right] \nn\\
\label{P+}
\eea
where ${\cal N}$ is a normalisation constant.  We used the symbol $g(v) = -\g v + F(v)$ for brevity. 
%
Reversing the velocities gives us the time reversed path  $X^\dagger=\{x'(t'), v'(t'), f'(t') \} =\{x(\t-t),-v(\t-t), f(\t-t)\}$, the probability of which can be expressed as
\bea
{\cal P}_- 
&=& {\cal N} \d(\dot x - v) \exp \left[ -\hf \int_0^\t dt  \f{\p g(v)} {\p v} \right] \nn\\
&\times& \exp\left[{-\frac{1}{4D_0} \int_0^\tau  dt \left( \dot v + g(v) + \f{\p U}{\p x} -f(t) \right)^2}\right] \nn\\
\eea
%
where in the last step it was assumed that $g(v)$ is an odd function $g(-v)=-g(v)$. 
This condition is naturally satisfied in many SPP models that assume energy transduction from internal energy source to kinetic energy~\cite{Romanczuk2012}, as will
be illustrated further in the following sections. 
Thus the ratio of the probabilities of the forward and reverse paths comes out to be
\bea
\f{{\cal P}_+}{{\cal P}_- } 
=\exp\left[{\frac{1}{D_0} \int_0^\tau  dt  \left( \dot v  + \f{\p U}{\p x} -f(t) \right)} g(v) \right]  \nn 
\eea

After some algebra one finally gets (see Appendix-\ref{ap_ratio}) 
\bea
\f{{\cal P}_+}{{\cal P}_- } 
 &=& \exp\left[ - \be \left(\D q + \D Q_{em} +\f{1}{\g} \D \phi \right) \right]
 \label{p+p-}
\eea
where $\be = 1/\kb T = \g/D_0$.  
In the above relation  $\D q$  is the heat flow identified in the first law.
The term $\D Q_{em} = (1/\g)\int_0^\t dt\,  F(v).(f(t)- \p_x U)$ is due to the coupling between the internal motor degrees of freedom and mechanical forces.  
$\D \phi$ is the change in a velocity dependent potential defined through $F(v) = -\p \phi(v)/\p v$.

Eq.(\ref{p+p-}) gives the ratio of the probabilities of forward and reverse paths, given that the forward evolution takes the system from  initial state $o$ to final state $e$. Assuming that the normalized probability distribution of these two states are $\pi_o$ and $\pi_e$ respectively, the ratio of the forward and the reverse processes is
\bea
\f{P_f(X)}{P_r(X^\dagger)}=\f{\pi_o{\cal P}_+}{\pi_e{\cal P}_- } &=& e^{\D s/\kb} e^{- \be \left(\D q + \D Q_{em} +\f{1}{\g} \D \phi \right)} \nn\\
&=& \exp[\D s_f/\kb] 
\label{pfpr}
\eea
where we used the {\em stochastic entropy} content corresponding to the distributions of initial and final {\em states} 
given by  $s_{o,e}=-\kb \ln \pi_{o,e}$~\cite{Seifert2005},  
leading to $\pi_o/\pi_e = \exp(\D s/\kb)$. 
The total entropy production in the forward process  is
\bea
\D s_f &=& \D s -\f{1}{T}\left(\D q + \D Q_{em} +\f{1}{\g} \D \phi \right) \nn\\
&=&  \D s -\f{1}{T}\left(\D E -\D W + \D Q_{em} +\f{1}{\g} \D \phi \right),
\label{s_tot}
\eea
where in the last step we used  Eq.(\ref{1st}).

The main contribution of this paper is the identification of this total entropy production Eq.(\ref{s_tot}) which contains two new terms as compared to a
system of traditional Brownian particles. 
These are the energy exchange between the motor's internal degrees of freedom and the external mechanical forces $\D Q_{em}$, 
and a change in the velocity dependent potential $\D \phi$. Both these contributions disappear once the motor activity of the self propelled particles is switched off.  
Note that both  $\D Q_{em}$ and $\D \phi(v)$ are hidden from the perspective of the first law, but appears in the expression of the total entropy change. 
This is due to the intrinsic {\em open} nature of the system with respect  to the self propulsion mechanism.

Eq.(\ref{pfpr}) implies  the integral fluctuation theorem~\cite{Kurchan2007}
\bea
\la e^{-\D s_f/\kb} \ra &=& \int {\cal D}[X] \f{P_r(X^\dagger)}{P_f(X)}P_f(X) \nn\\
 &=&  \int {\cal D}[X^\dagger] P_r(X^\dagger) =1.
 \label{ift}
\eea
This leads to $\la \D s_f \ra  \geq 0$,  a positive average entropy production.
Note that Eq.(\ref{ift}) together with Eq.(\ref{s_tot}) 
gives
$\la e^{-\be \D W} \ra = \la e^{-\be \D A} e^{-\be (\D Q_{em}+ \D \phi/\g)}\ra$
where $\D A=\D E-T \D s$. 
In the absence of motor driving, $\D Q_{em}=0$ and $\D \phi=0$, $\D A$ is the change in Helmholtz free energy, and the above relation gives the Jarzynski equation 
$\la \exp(-\be \D W)\ra = \exp(-\be \D A)$~\cite{Jarzynski1997}.



\subsection{ Stochastic thermodynamics at NESS} 
It can be shown that at NESS the detailed fluctuation theorem (see Appendix-\ref{ap_dft})
\bea
\f{\r(\D s_t)}{\r(-\D s_t)} = e^{\D s_t/\kb}
\label{dft}
\eea
holds, where $\D s_t$ is the total entropy change along any trajectory.

Calculation of total entropy takes a simple form when $U(x)=0$. We derive this result here,  as it will be used in the context of particular models of velocity dependent
force in later sections. The Langevin equation in presence of a constant external force is  
$\dot v = -\p \psi (v)/\p v + \eta$,
with $ -\p \psi (v)/\p v= -\g v + F(v) + f  $. The corresponding Fokker-Planck equation has the form
$\p_t p(v,t) = D_0 \p_v\left[  e^{-\psi/D_0} \p_v(e^{\psi/D_0}p)\right]$ 
with a steady state solution
\bea
p_s(v) =  \f{1}{Z}e^{-\psi(v)/D_0}
\label{ps0}
\eea
where the normalization $Z=\int_{-\infty}^\infty dv \exp(-\psi(v)/D_0)$, 
and $\psi(v) = (\g v^2/2 - f v) + \phi(v)$ with $\phi(v)=-\int dv F(v)$, as before. 
At NESS entropy is continually produced, and the system entropy change is $\D s/\kb = \D \psi/D_0$.
Thus the total entropy production (Eq.\ref{s_tot}) is
\bea
\f{\D s_t}{\kb} = \be \left(-\f{1}{\g} f\D v + \D W - \D Q_{em}\right),
\label{ds_ness}
\eea
where $-f \D v/\g$ denotes work done by the SPP due to its change in velocity. 
The energy flux $\D Q_{em}$  and work done $\D W$  have the same meaning as discussed in the previous subsection.  
The total entropy production in NESS obeys both the integral and detailed fluctuation theorems of Eq.s (\ref{ift})  and  (\ref{dft}).

\section{Linear response at NESS: modified fluctuation dissipation relation}
The Fokker-Planck equation corresponding to Eq.(\ref{lange})  is
\bea
\p_t p(x,v,t) = \cL(x,v,h) p(x,v,t) =(\cLo + f(t) \cL_1)p 
\eea
where
\bea
\cLo p &=& - \p_x(v p) - \p_v \left[ g(v) -\p_x U \right] p + D_0 \p_v^2 p \nn\\
\cL_1 p &=& -\p_v p. \nn
\eea
Assuming that the SPP system goes to a steady state characterized by a distribution function $p_s$ obeying  $\cLo p_s =0$, linear response
around this steady state is described by\cite{Chaudhuri2012, Seifert2010, Verley2011, Agarwal1972}
\bea
\f{\d \la A(t)\ra}{\d f(t')} = \la A(t) M(t') \ra_s
\eea 
where $\la \dots \ra_s$ indicate a steady state average, and 
$M = ({1}/{p_s}) \cL_1 p_s$.

The steady state distribution of Eq.(\ref{ps0}) leads to 
$M=\p_v [-\ln p_s] = \psi'(v)/D_0 = - g(v)/D_0 = (\g v -F(v))/D_0$. 
Therefore the response function is given by
\bea
\f{\d \la A(t)\ra}{\d f(t')} 
& =& \f{1}{D_0}\la A(t) [\g v(t') -F(v(t'))] \ra_s \nn\\
&=& \be \la A(t) v(t') \ra_s -  \f{1}{D_0} \la A(t) F(v(t')) \ra_s.
\eea
This is the modified fluctuation dissipation relation (MFDR) characterizing response function at NESS of SPP. 
In the absence of self propulsion,  $F(v)=0$, one gets back the equilibrium fluctuation dissipation theorem (FDT).
The velocity response to external force is
\bea
\chi(t,t')=\f{\d \la v(t)\ra}{\d f(t')} = \be \la v(t) v(t') \ra_s -  \f{1}{D_0} \la v(t) F(v(t')) \ra_s. \nn\\
\label{vfdt}
\eea
Since the correction in MFDR at NESS with respect to the equilibrium FDT is additive (see Eq.~\ref{vfdt}), not multiplicative, 
a ratio of the correlation and response $\la v(t) v(t') \ra/\chi(t,t')$ can not, in general, be interpreted as an effective temperature, with the 
only possible exception being when $F(v)$ is a linear function of velocity $v$.   

In the following, we consider some specific models of self propelled particles and analyze their behavior using the formalism developed so far.  
\section{models of spp} 
In this section we consider three specific examples of SPP. The first one is the simplest, uses a linear force-velocity relation that sometimes is used to describe Kinesin like motor proteins\cite{Howard2001}. The second and third example use non-linear velocity-dependent forces. The second example deals with the Rayleigh-Helmholtz model\cite{Romanczuk2012} which has been useful to describe the collective motion of a bunch of motor proteins working in tandem to move appropriate cargo\cite{Badoual2002}.  In the third example we use the energy-depot model~\cite{Schweitzer1998,Romanczuk2012}, which utilizes a simple coupling between internal energy production and mechanical motion to propel particles. 

\subsection{Molecular motors} 
Molecular motors, e.g., kinesins move on polymeric tracks, e.g., microtubules in a highly stochastic but directed manner 
utilizing chemical energy from ATP hydrolysis. In the presence of load force acting in the direction opposing their motion, 
they slow down and eventually stop moving. This behavior can be approximately modeled through a linear force-velocity relation~\cite{Svoboda1994}. 
Let us assume the autonomous force produced by the motor is $f_s$. In the presence of an external load force $-\l$, 
the Langevin equation is 
\bea
\dot v = -\g v + \eta + f_s -\l.
\label{lange_motor}
\eea
In the over-damped limit, this leads to the linear force-velocity relation $\la v \ra = v_0(1 - \l/f_s)$ with $v_0 = f_s/\g$ the autonomous velocity of free motors, and $f_s$ the stall force. 
Note that,  
using a linear velocity dependent force $f_s(1 - v/v_0)$ in place of $f_s$, merely changes the effective viscous drag $\g$ in the above equation by a constant additive amount.
In molecular motors, the mechano-chemical processes leading to self propulsion, in general, may elevate the noise level, change the noise correlation,
and change the viscous drag. However, in this simple model we assume that the noise can still be regarded as white if the time resolution is not too small, 
and the effective diffusion constant contains the impact of chemical reactions.    
In the absence of external load $\l=0$, the Langevin equation  can be rewritten as 
$\dot v = -\psi'(v) + \eta$ where
$\psi'(v)\equiv\p \psi/\p v=\g v -f_s$  is obtainable from 
$\psi(v) = (\g/2)(v-v_0)^2$. 
Thus the steady state distribution (Eq.~\ref{ps0})
\bea
p_s(v) = \sqrt{\f{\be}{2\pi}} \exp\left( -\f{\be}{2} (v-v_0)^2\right).
\label{ps_m}
\eea
%

\subsubsection{ Entropy production at NESS} 
The total entropy production can be obtained from Eq.(\ref{ds_ness}). 
Combining the constant self propulsion force $f_s$ with the load force $-\l$, the terms 
in Eq.(\ref{ds_ness}) $f=f_s-\l$, $F(v)=0$, $\D Q_{em}=0$ gives the total stochastic entropy production 
\bea
\f{\D s_t}{\kb} = \be \left[  -\f{1}{\g} (f_s-\l) \D v + \D W\right],
\eea
with $\D W = (f_s-\l) \int^\t v dt$.
The integral and the detailed  fluctuation theorems 
will be obeyed by this total entropy production at steady state. 

One can extend this calculation to rotating motors, by replacing linear displacements by rotation, 
velocities by angular velocities, and forces by torques.
Note that for measurements over asymptotically long time $\t$, $\D W$ in the above expression becomes predominant and 
hence $\D s_t/\kb = \be \D W$, the form used in recent experiments on F$_1$ATPase~\cite{Hayashi2010}.

\subsubsection{Entropy production at oscillatory steady states}
In the presence of a time-dependent external force the Langevin equation describing the molecular motor is
\bea
\dot v = -\g v + \eta + f_s + f(t),
\label{lange_m}
\eea
with the general solution at initial condition independent  asymptotic states 
$v(t) = v_0 + \int_0^t dt' e^{-\g (t-t')} [ f(t') + \eta(t')]$. 
Thus $v(t)$ is a linear functional of Gaussian noise $\eta(t')$, implying that the probability distribution of $v(t)$ is also Gaussian,
\bea
p(v,t) = \sqrt{\f{\be}{2\pi}} \exp\left( -\f{\be}{2 } (v-\la v(t) \ra)^2\right)
\eea 
where
$\la v(t) \ra =v_0+ \int_0^t dt' e^{-\g (t-t')} f(t')$. 
If the external force is sinusoidal $f(t) = A \sin \w t$,  the mean velocity at the asymptotic oscillatory state is
\bea
\la v(t) \ra = v_0 + \f{A}{\g^2 +\w^2} \left[  \w(1-\cos \w t) + \g \sin \w t  \right].
\eea
The system entropy production during a time $\t$ is 
$\D s/\kb = - \ln [p(v_\t,\t)/p(v_0,0)] = \be  (\bar v - \bar{\la v \ra}) (\D v -\D \la v\ra)$
with $\bar v  = ( v_\t +  v_0)/2$ and $\D  v =  v_\t -  v_0$. Thus the total entropy production is given by
\bea
\f{\D s_t}{\kb} = \be [ \D W  - \la \bar v \ra \D v - (\bar v - \la \bar v \ra) \D \la v \ra ].
\eea

\subsubsection{Linear response at NESS} 
It is straightforward to obtain the velocity response to a perturbing force around a steady state of  
free molecular motors using Eq.(\ref{vfdt}),
$\kb T {\d \la v(t)\ra}/{\d f(t')} =\la v(t) v(t') \ra_s - v_0^2$.
Using Eq.(\ref{lange_motor}) one can directly calculate the two-time correlation function
$\la v(t) v(t') \ra_s = v_0^2  + \kb T e^{-\g|t-t'|}$.
Thus one obtains the equilibrium-like response function
\bea
\f{\d \la v(t)\ra}{\d f(t')} = e^{-\g|t-t'|}.
\eea

In higher dimension, SPPs with a constant magnitude of self propulsion force $f_s$ can steer the direction of propulsion~\cite{Romanczuk2012}. 
Thus the impact of this force on the motion of SPP is different from an externally applied force which is constant both in magnitude and direction. 
However, in 1d this difference disappears within the model described by Eq.(\ref{lange_motor}). 
Steering the direction of self propulsion in 1d would mean switching the direction of motion from forward to backward.  
This is achieved in the following examples through non-linear velocity dependent self propulsion forces. 

\subsection{The Rayleigh-Helmholtz model} 

In the Rayleigh-Helmholtz (RH) model~\cite{Rayleigh1945} one assumes a non-linear velocity dependent force
$F(v)=a v - b v^3$. This is sometimes interpreted as a viscous force $F(v)=-\g_1(v)v$ with a viscosity 
$\g_1(v)=-a+b v^2$ where $-a$ acts like a negative friction that pumps energy into the system. 
In the deterministic limit, this model has two fixed points at $v=\pm \sqrt{a/b}$.
In presence of a Langevin heat bath characterised by a viscous drag $\g$, the SPPs within RH model will experience a net negative 
drag $\g' = \g-a$ if $a > \g$, and the  stochastic noise can switch the particles between positive and negative velocities $\pm \sqrt{(a-\g)/b}$. 
%
The RH model has recently been 
used in various studies of SPPs~\cite{Erdmann2000, Strefler2008, Lindner2007, Romanczuk2012}, and 
describes the bimodal velocity distribution of microtubules under the collective influence of  bidirectional motor proteins 
NK11~\cite{Badoual2002}.  

\subsubsection{Entropy production at NESS}
In the presence of a constant external force $f$, one can write the total deterministic force  as $-\psi'(v) = -\g v + F(v) + f$ such that
$\psi(v) 
= (\g/2) v^2 - (a/2) (v-v_f)^2 + (b/4) v^4 $ 
with $v_f = (f/a)$. 
Thus, using Eq.(\ref{ps0}) one can find the steady state distribution
\bea
p_s(v,f) = \f{1}{Z} \exp\left[-\be \left( \f{v^2}{2} -\f{\a}{2} (v-v_f)^2 + \f{\nu}{4} v^4\right) \right]
\label{ps_rh1}
\eea
where $ \a=a/\g$ and $\nu=b/\g$. 
The corresponding stochastic entropy content is $s = -\kb \ln p_s$. 
%
It is straightforward to use Eq.(\ref{ds_ness}) to obtain the total entropy production within a NESS.
In a transformation from an initial state $p_s(v_0, f)$ to a final state $p_s(v_\t,f)$,
\bea
\f{\D s_t}{\kb} = \be \left[  -\a v_f \left( v_\t - \f{v_f}{2} \right) - (\a-1)\D W +  \D W_0\right],
\eea
where $\D W = f \int^\t v dt$ and 
$\D W_0=  \nu \int^\t v^3 dt$ grow with time $\t$ and are the asymptotically dominant terms. 

\subsubsection{Linear response at NESS}
The modified fluctuation dissipation relation, in this case, has the form
\bea
\chi(t,t')
&=& \be \la v(t) v(t') \ra_s - \f{1}{D_0} \la v(t) [av(t') - b v^3(t')] \ra_s \nn\\
&=& -\be(\a-1) \la v(t) v(t') \ra_s + \be \nu  \la v(t) v^3(t') \ra_s.
\eea
Note that at  $\a=0=\nu$ we have equilibrium, and obtain a fluctuation dissipation ratio $ \la v(t) v(t') \ra_s/\chi(t,t')=\kb T$. 
However, in general this ratio  depends on higher order correlations, and therefore on other quantities characterising a steady state. 

\subsubsection{Entropy production: external harmonic trap}
If an initially free SPP is subjected to an external harmonic potential $\hf k x^2$, the initial steady state described by Eq.(\ref{ps_rh1})
undergoes transformation to a final steady state achieved in the trapping potential. 
The  Langevin equation describing the dynamics of the SPP in trap is,
\begin{equation}
\dot{v}=-\g v  + \eta (t) + F(v) - k x. \nn
\end{equation}
The harmonic trap couples the time evolution of velocity with position.
Multiplying the above equation by $v$ we get a Langevin equation for the time evolution of the Hamiltonian
$\frac{dH}{dt}  = (g(v)+\eta)v $ 
where $H=v^2/2 + k x^2/2$, and $g(v)=-\g v + F(v)$ as before. 
In the deterministic limit of $\eta=0$, the motion goes to fixed points governed by $g(v)=0$ at $ v= \pm v_0$  with $v_0 = \sqrt{(a-\g)/b}$. 
This dynamics is characterised by $x=x_0 \sin (\w t + \phi)$, $v=v_0 \cos(\w t +\phi)$ with $\w^2=k$ and $x_0=v_0/\w$. The 
corresponding energy near these fixed points is $H \simeq H_0 = v_0^2$. 
The stochastic dynamics around these fixed points is described by the following Langevin equation~\cite{Schimansky-Geier2005} 
 ${dH}/{dt}=-\g_H H + \sqrt H \eta $
where $\g_H = \g -a + b H$. 
The corresponding Fokker-Planck equation
\begin{equation}
\frac{\partial P_H(H,t)}{\partial t}= \f{\p}{\p H} \left[ (\g _H H -D_0) P_H + D_0 \f{\p}{\p H}(H P_H)  \right] \nn
\end{equation}
has the  steady state solution
\begin{equation}
p_s (H) = \mathcal{A} \exp \left[-\frac{1}{D_0}\int \gamma _H  dH\right].\nn
\end{equation}
The most probable energy is given by the fixed point $H=\hf v^2+ \hf k x^2 = H_0$ where $\g_H(H_0)=0$.  
Thus near $H=H_0$ we can expand $\g_H$  as
$\gamma _H =b(H-H_0)$
to obtain
$p_s (H) =\mathcal{A} \exp \left[-\frac{b}{2D_0} (H-H_0)^2\right]$ 
which is equivalent to
\bea
\tilde p_s (x,v) &=& \mathcal{B}  e^{-\frac{\be \nu}{2} \left(\f{v^4}{4}-H_0 v^2\right)} \nn\\
&& e^{-\frac{\be \nu}{2 } \left(  \f{1}{4}\omega ^4 x^4 - H_0 \omega ^2 x^2\right)} 
e^{-\frac{\be \nu}{2} \omega^2 x^2 v^2}.
\label{ps_rh2}
\eea
Note that the 
term $\exp[-(\be\nu \w^2/2)\, x^2 v^2 ]$ implies that the particles with 
higher kinetic energies tend to locate near the potential minimum. 

We now determine the change in 
entropy as the initial steady state characterized by $p_s(v_i,0)$ (Eq.\ref{ps_rh1}) is transformed
to $\tilde p_s (x,v_f)$ given by Eq.(\ref{ps_rh2}). The change in system entropy is $\D s/\kb = -\ln[\tilde p_s(x,v_f)/p_s(v_i,0)]$, and 
the total entropy production in a trajectory (Eq.\ref{s_tot}),
\bea
\f{\D s_t}{\kb}
&=&  \f{\be}{2} \left[ -\f{\nu}{4} (v_f^4 + \w^4 x^4) - (\a-1) \w^2x^2 + \nu \w^2 x^2 v_f^2 \right]  \nn\\
&-& \be [ (\a-1) \D W -  \D W_0  ]
\eea
follows the integral and detailed fluctuation theorems given by 
Eq.s (\ref{ift}) and (\ref{dft}).

\subsection{The energy depot model}
Within the energy depot model~\cite{Schweitzer1998}, an SPP is capable of taking up external energy and store it in the internal energy depot, then transduce the energy
into kinetic energy. A part of the stored energy is dissipated during conversion into kinetic energy.
Thus the energy balance equation for an internal energy  $e(t)$ is 
$de(t)/{dt}=q(\textbf{r})-c e(t) -h(\textbf{v}) e(t)$
where $q(\textbf{r})$ is the space dependent rate of energy uptake, and $h(\textbf{v})$ is the rate of conversion of internal energy to kinetic energy.
In a particular simple version of the model, one makes the choice
$q(\textbf{r})=q_0$, i.e., uniform energy uptake and $h(\textbf{v})=d\textbf{v}^2$, conversion rate proportional to the kinetic energy itself. 
%
Assuming that  $e(t)$  reaches its steady state value at a much shorter time scale than the particle diffusion time, we use its steady state value
$
e_0=\frac{q_0}{c+d\textbf{v} ^2}.
$
Then the self propulsion force  is given by,
\begin{equation}
F(v)=ae_0v=\frac{aq_0v}{c+dv^2}.
\label{fv_edepot}
\end{equation}
In the limit of small velocities this model reduces to the Rayleigh-Helmholtz model $F(v)=\g_1 v - \g_2 v^3$ where $\g_1=a q_0/c$ and $\g_2=a q_0 d/c^2$.
Writing the total deterministic force acting on the SPP $-\psi'(v) = -\g v + f + a q_0 v/(c+d v^2)$, one gets 
$\psi(v) = \hf \g v^2 - f v - ({a q_0}/{2 d}) \ln\left( c + d v^2\right)$.
Thus using Eq.\ref{ps0}, the steady state distribution is 
\bea
p_s (v,f) 
 &=& \f{1}{Z} (c + d v^2)^{a q_0/2 D_0 d} \,\,e^{-\f{\be}{2}v^2 + \f{f v}{D_0}}.  
 \eea

The steady state entropy production due to a transformation from initial state $p_s(v_0, f)$ to a final state 
$p_s(v_\t,f)$ is given by Eq.\ref{ds_ness}, with $\D W = f \int^\t dt v$, $\D v = v_\t-v_0$ and 
$\D Q_{em}=(f/\g)\int^\t dt v/(c+ d v^2)$. 

The linear response around $f=0$ steady state can be expressed in terms of the modified fluctuation dissipation relation Eq.(\ref{vfdt})
with $F(v)$ given by Eq.(\ref{fv_edepot}), leading to
\bea
\chi(t,t')  = \be \la v(t) v(t') \ra -  \f{aq_0}{D_0} \left \la v(t) \f{v(t')}{c + d\, v^2(t')}\right \ra. 
\eea


\section{conclusion}
We have presented a stochastic thermodynamic description of non-interacting self propelled particles in terms of energy conservation and fluctuation theorem. 
This enabled us to identify the components of stochastic entropy production associated with non-equilibrium processes in SPPs. 
%
%
%
We studied entropy production for a simple model of molecular motors, the Rayleigh-Helmholtz model, and the energy depot model.
Calculation of entropy production and fluctuation theorems in SPPs  has become important in view of recent experimental 
interest in measurement of force generation by molecular motors using the detailed fluctuation theorem~\cite{Hayashi2010, Hayashi2012}.  

We further characterized the steady state response function in terms of a modified fluctuation-dissipation relation.
This in general has an additive correction due to self propulsion, compared to the fluctuation-dissipation theorem at equilibrium. 
Our predictions for the Rayleigh-Helmholtz model are particularly amenable to experimental verification,
due to its close relation to the motion of microtubules under collective influence of 
molecular motors NK11~\cite{Badoual2002,Endow2000},   
%

\acknowledgements
DC thanks Sriram Ramaswamy, Arnab Saha,  Abhishek Chaudhuri and  Arya Paul  for valuable discussions and critical comments on the manuscript. 
DC also thanks MPIPKS, Dresden for hospitality, where a part of this work was carried out.

\appendix
\section{Forward path probability}
\label{ap_fpath}
The probability of forward process is governed by the distribution of noise $\eta$ over time $\t$~\cite{Narayan2004}
\bea
P[\eta] = {\cal N} e^{-\f{1}{4D_0} \int_0^\tau  dt \eta^2(t)}
\eea
in presence of constraints
$\dot v = g(v) - \p U/\p x + f(t) + \eta(t) $ and $\dot x =v$. 
With paths denoted by $X=\{x(t),v(t),f(t)\}$, the forward path probability~\cite{Seifert2008,Imparato2006,Mallick2011} ${\cal P}_+[X] = J\, P[\eta]$ 
where 
%
the Jacobian $J={\rm det}{\bf M}$ and
the Jacobi operator 
\beq
{\bf M} = \f{\p \eta}{\p v} = \f{\p}{\p t} - \f{\p g(v)}{\p v}.
\eeq
The Jacobian can be written as~\cite{Mallick2011} 
\beq
J = \exp[ \mbox {Tr}(\ln {\bf M})] = \exp\left[ -\hf \int_0^\t dt \f{\p g}{\p v} \right]
\eeq
where in obtaining the last step, we discretised $\bf M$ and used Stratonovich convention (see Appendix-A of Ref.\cite{Seifert2008}).   

The discretisation process can be made explicit as follows. 
The time evolution of velocity can be discretised 
using the Stratonovich mid-point rule 
\bea
\f{v_i - v_{i-1}}{\e} &=& \hf [ g_i(v_i) + g_{i-1}(v_{i-1}) ] - \hf \left[\f{\p U}{\p x_i} +\f{\p U}{\p x_{i-1}} \right] \nn\\
&&  +\hf[ f_i + f_{i-1}] + \eta_i
\eea
where $\la \eta_i \ra=0$ and $\la \eta_i \eta_j \ra = 2 (D_0/\e) \d_{ij}$ with total time $\t=N \e$ discretised in $N$ equal steps of size $\e$.
Thus in the discretised notation the $ik$-th element of the $N \times N$ Jacobi matrix $\bf M$ is 
\bea
M_{ik} &=& \f{\p \eta_i}{\p v_k} = \f{1}{\e}(\d_{i,k}-\d_{i-1,k}) \nn\\
&& - \hf [g'_i(v_i) \d_{i,k} + g'_{i-1}(v_{i-1})\d_{i-1,k}]
\eea
where $g'_i(v_i)=\p g_i(v_i)/\p v_i$. Evaluating the determinant of this matrix one finds the Jacobian
\bea
J &=& {\rm det} {\bf M}=\left( \f{1}{\e}\right)^N \prod_{i=1}^N \left( 1 -\f{\e}{2} g'_i(v_i)\right)\nn\\
&=& \left( \f{1}{\e}\right)^N \exp \left[ \sum_{i=1}^N \ln \left(1-\f{\e}{2} g'_i(v_i) \right) \right] \nn\\
&\approx& \left( \f{1}{\e}\right)^N  \exp \left[-\sum_{i=1}^N \f{\e}{2} g'_i(v_i) \right] 
\eea
Apart from a multiplicative constant, the Jacobian in the continuum limit can be expressed as,
\bea
J = \exp \left[ - \hf \int_0^\t dt \f{\p g}{\p v} \right]. 
\eea

Using the constraints of equations of motion, thus, one can obtain the probability of forward path~\cite{Imparato2006}
\bea
{\cal P_+} &=& \d(\dot x -v) {\cal N} e^{  -\f{1}{4D_0} \int_0^\tau  dt \left( \dot v - g(v) + \f{\p U}{\p x} -f(t) \right)^2} 
e^{ -\hf \int_0^\t dt \f{\p g}{\p v} }. \nn\\ 
\eea

\section{Ratio of probabilities}
\label{ap_ratio}
The ratio of the probabilities of the forward and reverse paths comes out to be
\bea
\f{{\cal P}_+}{{\cal P}_- } 
&=& \exp\left[{\frac{1}{D_0} \int_0^\tau  dt  \left( \dot v  + \f{\p U}{\p x} -f(t) \right)} g(v) \right] \nn\\
 &=& \exp\left[ \frac{1}{D_0} \int_0^\tau  dt  \left( -\g v +\eta + F(v) \right) (-\g v +F(v) ) \right] \nn\\
\eea
where in the last step we used the Langevin equation and the expression of $g(v)$.
The terms  in the exponential  can be rewritten in the form
\bea  
 && \int_0^\t      dt (-\g v+\eta)  (-\g v) + F(v) (F(v) -2 \g v + \eta)   \nn\\
&=& -\g \D Q + {\cal I}
\eea
where the definition of $\D Q$ is used from the first law Eq.(\ref{1st}),
with the second term
\bea
{\cal I} &=& \int_0^\t dt F(v) [ \dot v - \g v - (f(t)- \p_x U)] \nn\\
&=&  -\int_0^\t dt \left[  \dot v  \f{\p \phi}{\p v} + \g v.F(v) + F(v).(f(t)- \p_x U) \right] \nn\\
& = & -\D \phi - \g \D Q_m - \g \D Q_{em}.
\eea
In the first term in the expression of ${\cal I}$, we have used $F(v) = -\p \phi(v)/\p v$ and thus
$\int_0^\t dt\, \dot v.\p_v\phi(v) = \phi(v(\t)) - \phi(v(0)) = \D \phi $, the change in velocity dependent potential.
The second term $\D Q_m = \int^\t dt v.F(v)$ is defined in the first law. In the third term we used the definition 
$\g \D Q_{em} =   \int^\t dt F(v).(f(t)- \p_x U)$. Thus we get the ratio in Eq.(\ref{p+p-}).

\section{Detailed fluctuation theorem}
\label{ap_dft}
It follows from Eq.(\ref{pfpr}) that the probability distribution of entropy production~\cite{Crooks1999,Kurchan2007}
\bea
\r(\D s_t) &=& \int {\cal D}[X] P_f(X) \d( \D s_t - \D s_f(X)) \nn\\
&=& \int {\cal D}[X] P_r(X^\dagger) e^{\D s_f/\kb}   \d( \D s_t - \D s_f(X)) \nn\\
&=&  e^{\D s_t/\kb}  \int {\cal D}[X^\dagger] P_r(X^\dagger)  \d( \D s_t + \D s_r(X^\dagger))  \nn\\
&=& e^{\D s_t/\kb}  \r(-\D s_t)
\eea
where we used 
$\D s_f(X) = -\D s_r(X^\dagger)$, 
i.e., the final distribution of the forward process is assumed to be
the same as the initial distribution of the reverse process, and vice versa~\cite{Crooks1999}. This assumption is valid at steady states. 

\bibliographystyle{apsrev4-1}

%

\end{document}